\documentclass[11pt,preprint]{aastex}
\usepackage{natbib}
\begin{document}
\shorttitle{ALMA Observations of the Orion Proplyds}
\shortauthors{Mann et al.}

\title{ALMA observations of the Orion Proplyds}
 
 \author{Rita K. Mann\altaffilmark{1}, James Di Francesco\altaffilmark{1,2}, Doug Johnstone\altaffilmark{1,2}, 
 Sean M. Andrews\altaffilmark{3}, Jonathan P. Williams\altaffilmark{4}, John Bally\altaffilmark{5}, 
 Luca Ricci\altaffilmark{6}, A. Meredith Hughes\altaffilmark{7}, Brenda C. Matthews\altaffilmark{1,2}}

\altaffiltext{1}{National Research Council Canada, 5071 West Saanich Road, Victoria, BC, V9E 2E7, Canada} 
\altaffiltext{2}{Department of Physics and Astronomy, University of Victoria, Victoria, BC, V8P 1A1, Canada } 
\altaffiltext{3}{Joint Astronomy Centre, 660 North A'ohoku Place, University Park, Hilo, HI 96720, USA}
 \altaffiltext{4}{Harvard-Smithsonian Center for Astrophysics, 60 Garden Street, Cambridge, MA 02138, USA}
 \altaffiltext{5}{Institute for Astronomy, University of Hawaii, 2680 Woodlawn Drive, Honolulu, HI 96822 USA}
 \altaffiltext{6}{CASA, University of Colorado, CB 389, Boulder, CO 80309, USA}   
 \altaffiltext{7}{Department of Astronomy - California Institute of Technology, MC 249-17, Pasadena, CA 91125, USA}
 \altaffiltext{8}{Van Vleck Observatory, Astronomy Department, Wesleyan University, 96 Foss Hill Drive, Middletown, CT 06459, USA }

\email{rita.mann@nrc-cnrc.gc.ca}

\begin{abstract}
We present ALMA observations of protoplanetary 
disks (``proplyds'') in the Orion Nebula Cluster.  We imaged 5 individual 
fields at 856\,$\micron$ containing 22 {\em HST}-identified proplyds and 
detected 21 of them. Eight of those disks were detected for the first time 
at submillimeter wavelengths, including the most prominent, well-known 
proplyd in the entire Orion Nebula, 114-426.  Thermal dust emission in 
excess of any free-free component was measured in all but one of the 
detected disks, and ranged between 1--163\,mJy, with resulting disk 
masses of $0.3-79\,M_{jup}$.  An additional 26 stars with no prior evidence 
of associated disks in {\em HST} observations were also imaged within 
the 5 fields, but only 2 were detected.  The disk mass upper limits for 
the undetected targets, which include OB stars, $\theta^1$\,Ori C and 
$\theta^1$\,Ori F, range from 0.1--0.6\,$M_{jup}$.  Combining these 
ALMA data with previous SMA observations, we find a lack of massive 
($\gtrsim$\,3\,$M_{\rm jup}$) disks in the extreme-UV dominated region 
of Orion, within 0.03\,pc of $\theta^1$\,Ori C.  At larger separations from 
$\theta^1$\,Ori C, in the far-UV dominated region, there is a wide range 
of disk masses, similar to what is found in low-mass star forming regions.  
Taken together, these results suggest that a rapid dissipation of disk 
masses likely inhibits potential planet formation in the extreme-UV 
dominated regions of OB associations,
but leaves disks in the far-UV dominated regions relatively unaffected.

\end{abstract}


\section{Introduction}\label{sec: intro}
Circumstellar disks are the birthsites of exoplanets.  
{\it Hubble Space Telescope} ({\it HST}) images of the Orion Nebula Cluster (ONC)
revealed a hostile environment; many of the disks that orbit low mass 
stars are being photoevaporated by the intense UV radiation from the most 
massive nearby star, $\theta^1$\,Ori C \citep[spectral type O6;][]{odell94,
mccullough,bally98,smith,ricci}.  These disks are surrounded by tear-drop 
shaped structures, with bright heads facing $\theta^1$\,Ori C and tails 
pointing radially away.  These distinctive circumstellar morphologies 
led to the nomenclature ``proplyds", an acronym for PROtoPLanetarY 
DiskS, that is now regularly applied to low-mass stars and their disks in 
the centers of massive star forming regions \citep{odell94}. 
The Orion proplyds were found to suffer photoevaporative mass-loss rates of 
$\dot{M} \approx 10^{-7}$\,$M_\odot$\,yr$^{-1}$ \citep{churchwell,henney}, 
high enough to disperse the amount of disk mass required to form a 
planetary system like our own in under 1\,Myr.  That dissipation timescale 
is too short compared to the core accretion requirements for giant planet 
formation \citep[e.g.,][]{hubickyj05}, and is in apparent conflict with the 
inferred ages of the ONC stars \citep[$\sim$2\,Myr;][]{reggiani11,dario}.

Measurements of the masses that remain in the Orion proplyds are crucial for characterizing
the photoevaporation process and assessing their potential for planet formation.
The most straightforward way to estimate a disk mass is from 
a measurement of the thermal dust continuum luminosity at long 
wavelengths, where the emission is optically thin \citep[cf.][]{beckwith90}.  
Although molecular gas likely comprises the vast majority of the mass budget in 
disks, the dust dominates the opacity and is significantly easier to detect.  
The Berkeley-Illinois-Maryland Array (BIMA), Owens Valley Radio Observatory 
(OVRO), Plateau de Bure Interferometer (PdBI), and Combined Array for Research 
in Millimeter Astronomy (CARMA) observed the Orion proplyds at wavelengths of 
1.3--3.5\,mm \citep{mundy,bally98b,elada98,eisner06,eisner08}, but 
unfortunately these data provided limited constraints on the disk masses: 
contamination by free-free radiation from the ionized cocoons generated by the 
photoevaporation process often dominated the dust emission.  Soon after the 
Submillimeter Array (SMA) was commissioned, it produced the first successful 
detections of the Orion proplyds at a submillimeter 
wavelength (880\,$\mu$m), where the dust emission dominates \citep{williams}.  Those 
observations revealed that at least some of these disks still have sufficient 
mass ($>$ 10\,M$_{\rm jup}$) remaining to potentially form giant planets.  
A larger scale SMA survey of 
the Orion proplyds identified the erosion of the high end of the disk mass 
distribution due to photoevaporation by $\theta^1$\,Ori C \citep{mann09a,
mann10}.  To date, however, these surveys have only been sensitive enough to 
detect the most massive disks ($\ge$\,8.4\,M$_{\rm jup}$) in the ONC, and 
therefore provide relatively biased information about disk evolution in the 
hearts of massive star forming regions.  

To probe the full disk mass distribution in the ONC and further study the impact of 
external photoionizing radiation on disk properties, we carried out a much more 
sensitive pilot survey with the Atacama Large Millimeter/submillimeter Array 
(ALMA) that targeted 48 young stars, including 22 {\it HST}-identified 
proplyds.  In this article, we present the results of the 856\,$\mu$m (ALMA 
Band 7) continuum observations.  The observations and data reduction are 
described in \S \ref{obs}.  Estimates of disk masses are presented in \S 
\ref{results}, and an examination of the dependence of disk mass on location in 
the ONC is discussed in \S \ref{disc}. 

\section{Observations}\label{obs}
\begin{figure}[h]
\centering
\includegraphics[scale=0.7]{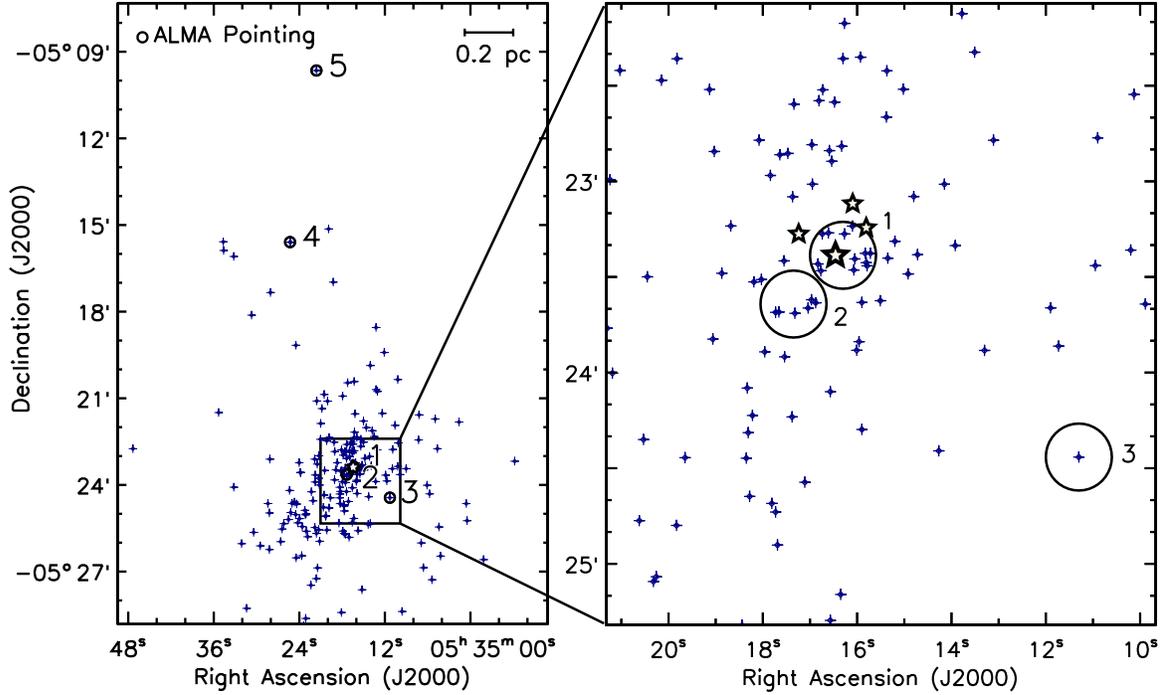}
\vskip 0in
\caption[Figure1]{
Location of the 5 observed ALMA fields in the Orion Nebula 
cluster. The stars mark the position of the OB stars, and blue 
crosses show the location of the proplyds identified by 
{\em HST} observations \citep{ricci}. 
Black circles represent the 18\arcsec\ primary beam 
of the ALMA observations at 856\,\micron.
Fields 1-5 are labeled according to Table \ref{table1}.  
The black square outlined in the left panel is zoomed in 
for the panel on the right to allow a better view of the 
crowded central fields, 1-3, near the OB stars.
Fields 1-3 contain 42 of the stars, while Fields 4 \& 5 
contain the remaining five young stars.}
\label{fig1}
\end{figure}

Five individual pointings (hereafter Fields) in the ONC were observed with 
ALMA using the Band 7 (345\,GHz) receivers on 2012 October 24, as part of the 
Cycle 0 Early Science operations (see Table \ref{table1}).  Figure \ref{fig1} 
marks the pointing centers, with reference to the high mass members of the 
ONC.  Twenty-two 12-m diameter 
ALMA antennas were arranged in a hybrid configuration that yielded (with 
robust weighting) images with a $\sim$0$\farcs$5 angular resolution.  This 
target resolution was chosen to distinguish individual disks toward the crowded 
Trapezium cluster, resolve the emission from three large proplyds, and filter
out potentially confusing large-scale emission from the background molecular 
cloud.  Each Field was observed for 136 seconds per visit.  Fields 2-5 were 
visited six times over 6.5 hours to improve sampling in the Fourier plane, but 
Field 1 was observed twice as often as the others to achieve higher sensitivity 
for the disks nearest to $\theta^1$\, Ori C.  The correlator was configured to 
observe simultaneously four 1.875\,GHz-wide spectral windows, each divided into 
3840 channels each with a width of 488.28\,kHz; after online Hanning smoothing, 
the spectral resolution was 976.56 kHz.  The spectral windows were arranged 
to cover the CO\,(3--2), HCN\,(4--3), HCO$^+$\,(4--3) and CS\,(7--6) emission 
lines.  The focus here will be on the wideband ($\Delta \nu \approx7.5$\,GHz) 
continuum emission, extracted by integrating over all line-free channels: the 
spectral line data will be presented elsewhere.  At 856\,$\mu$m, the mean 
wavelength of the four spectral windows, the effective field of view is 
18\arcsec\ (the FWHM primary beam of an individual antenna).  Since the
shortest baseline was 21.2 m in length, the maximum recoverable scale was
4$\farcs$99.

Data calibration and image reconstruction were performed using standard 
procedures in the {\tt CASA} package.  The antenna-based complex gains were 
calibrated based on repeated observations of the quasar J0607--085.  The 
absolute amplitude scale was determined from observations of Callisto, and the 
bandpass response of the system was measured from observations of the bright 
quasar J0522--364.  The model of Callisto was that provided by Butler (2012).
The mean and standard deviation of the 856\,$\mu$m continuum flux of Callisto 
over all five scheduling blocks and all line-free channels in four spectral 
windows was 18.52 Jy and 0.53 Jy, respectively.  Each field was Fourier-inverted 
separately, using the multi-frequency synthesis mode in the {\tt CLEAN} task 
with an intermediate Briggs robustness parameter of 0.5 chosen to achieve the 
desired angular resolution of $\sim$0$\farcs$5.  A dirty image was generated 
out to a diameter of $\sim$34\arcsec\ (only $\sim$2\%\ of the maximum primary 
beam sensitivity), in an effort to include any bright disks that might have 
been located at the far margins of each field.  These dirty images were then 
{\tt CLEAN}ed with the Clark algorithm down to a threshold of $\sim$2$\times$ 
the RMS observed in emission-free locations near the pointing centers (boxes 
were used to focus the algorithm toward features that could be visually confirmed 
in the dirty images).  After {\tt CLEAN}ing, each field was corrected for primary 
beam attenuation and restored with a synthesized 0$\farcs$51 $\times$ 0$\farcs$46 
beam.  The 1 $\sigma$ rms levels obtained toward the center of each Field are 
listed in Table \ref{table1}.

\section{Results}\label{results}

\begin{figure}[h]
\centering
\includegraphics[scale=1.6]{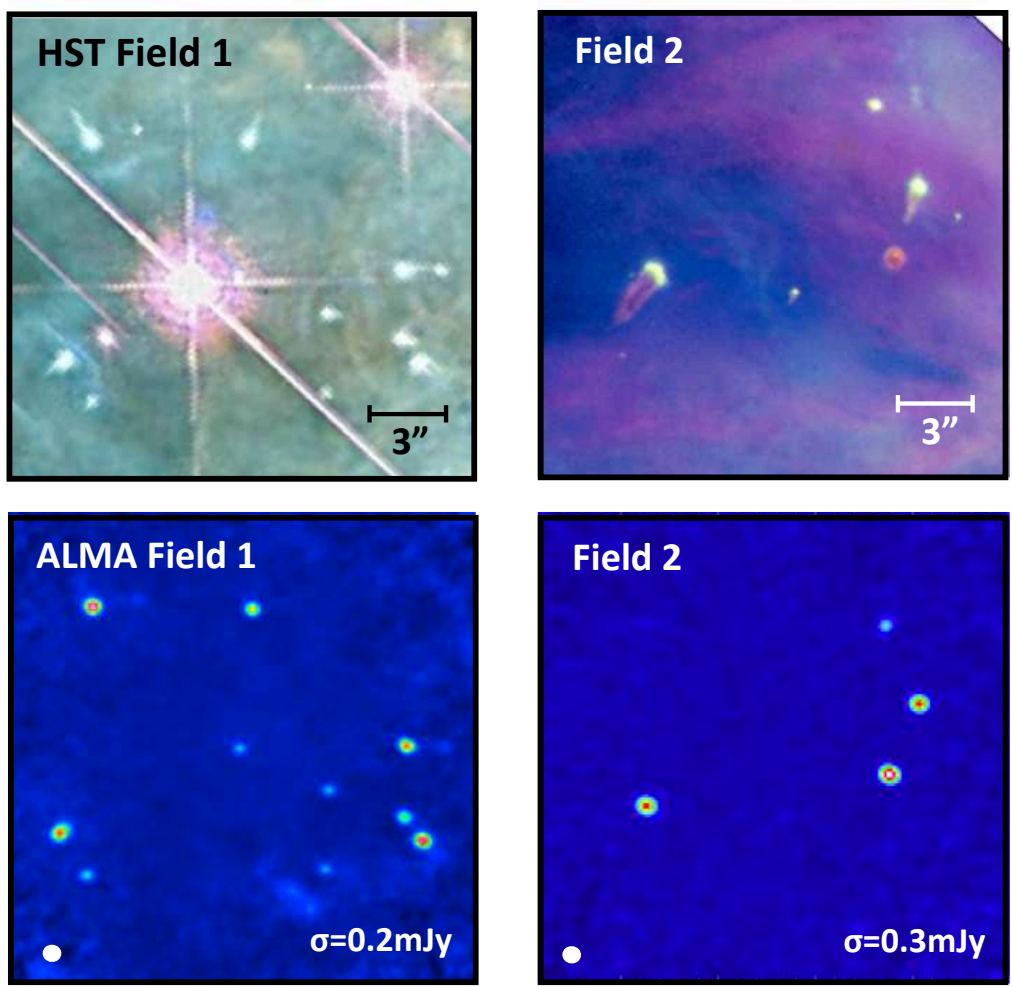}
\vskip -0.0in
\caption[HST+ALMA]{
Top: {\em HST} images from \citet{bally00} of Fields 
1 and 2 that were observed with ALMA in Cycle 0.
Bottom: Corresponding ALMA 0$\farcs$5 FWHM 
resolution 856$\micron$ observations.
These fields have image sizes of 
20$\arcsec\times$20$\arcsec$ and are centered
at the positions shown in Table \ref{table1}.
The ALMA images are the highest resolution 
observations taken at submillimeter wavelengths
of the central regions of the Orion Nebula cluster.}
\label{fig2a}
\end{figure}

\begin{figure}[h]
\centering
\includegraphics[scale=1.2]{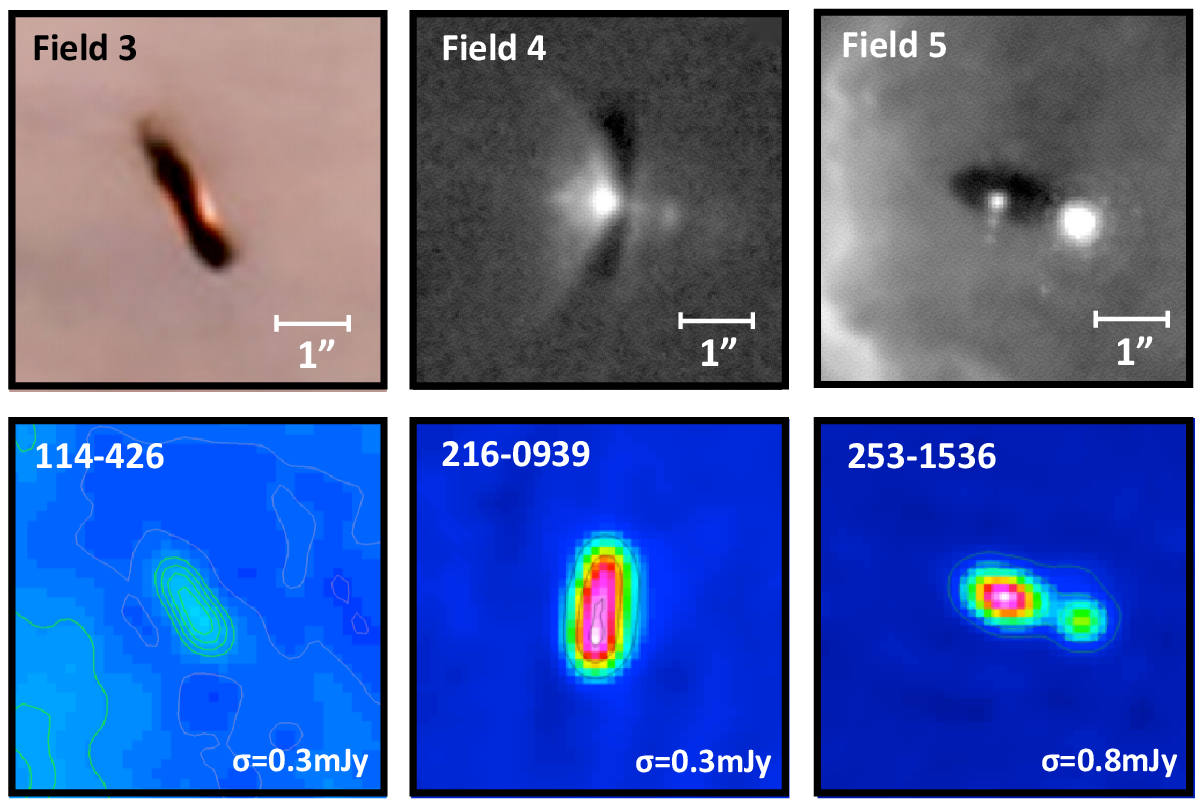}
\vskip -0in
\caption[HST+ALMA2]{
Top: {\em HST} images from \citet{bally00} and \citet{smith}
of Fields 3-5 observed with ALMA in Cycle 0.
Bottom: Corresponding ALMA 0$\farcs$5 
FWHM resolution 856$\micron$ observations.
The image sizes are 5$\arcsec\times$5$\arcsec$.}
\label{fig2b}
\end{figure}

The 856\,$\mu$m continuum maps of the surveyed ONC fields are 
shown in Figures \ref{fig2a} and \ref{fig2b}, alongside the corresponding 
optical {\em HST} images \citep{bally00,smith}.  
These ALMA images represent the highest resolution observations at 
submillimeter wavelengths towards the central OB stars in the ONC.
We detected submillimeter emission from 21 of the 22 targeted
{\it HST}-identified proplyds in this survey; the only proplyd not detected 
is 169--338 (see Table \ref{table2}). 
Eight of those detections are 
new, and 13 are recoveries of previous SMA detections.  
The centroid positions and 856\,$\mu$m integrated flux densities 
for each target were measured by fitting elliptical Gaussians in the image 
plane.  A suitable RMS noise level in each field was 
determined from the emission-free regions within the primary beam.

\begin{figure}[h]
\centering
\includegraphics[scale=0.5]{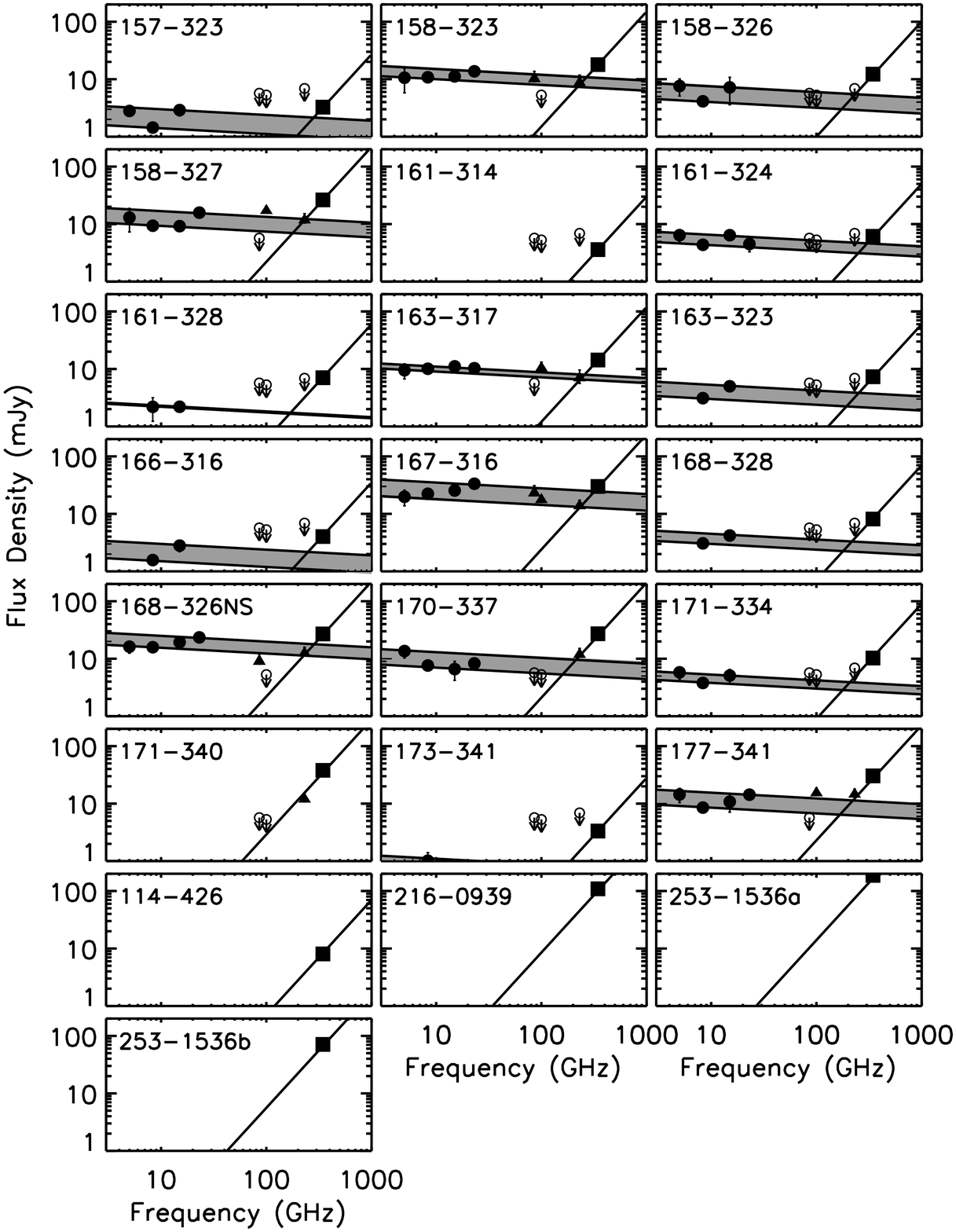}
\caption[SED]{
Radio-submillimeter spectral energy distributions of the Orion 
proplyds detected at $\geq 3\sigma$ with ALMA at $856\,\mu$m.
The ALMA measurements are represented by squares, 
centimeter observations by circles \citep{garay,felli93,zapata},
and millimeter observations by triangles \citep{mundy,bally98b,eisner06,eisner08}.
Open circles are upper limits from non-detections and
uncertainties not shown are smaller than symbol sizes.
The extrapolated range of optically thin free-free emission,
$F_\nu \propto \nu^{-0.1}$, is overlaid in gray.
A template to the disk emission, $F_\nu \propto \nu^2$, is
shown to guide the eye and reveal the
relative contribution of the ionized gas and dust components.
}
\label{sed}
\end{figure}

The observed emission is composed of a free-free ($F_{\rm ff}$) 
contribution from the ionized 
cocoons that surround the photoevaporating disks and the thermal 
dust emission ($F_{\rm dust}$) from the disks themselves, such that $F_{\rm 
obs} = F_{\rm dust} + F_{\rm ff}$.  
The radio-submillimeter spectral energy distributions (SEDs) for the 22 disks
detected at 3$\sigma$ at 856\,$\mu$m with ALMA are shown in Figure \ref{sed}.
The free-free contributions from the disk targets were extrapolated from 
centimeter wavelengths into the submillimeter regime using published 
VLA flux densities from 1.3\,cm to 6\,cm \citep{garay,felli,felli93,zapata}.  
Fits to the free-free emission ($F_{\rm ff} \propto \nu^{-0.1}$)
and dust emission ($F_{\rm dust} \propto \nu^{2}$) are overlaid on the SEDs to show
their relative contributions and contrasting spectral dependences. 
The radio observations show the flat spectral dependence consistent
with optically thin emission, but with a range, highlighted by the grey
scale, which we attribute to variability \citep{felli93,zapata}.
We avoided observations taken at wavelengths longer than 6\,cm (5 GHz) in
this analysis, in order to avoid the turnover frequency, 
where the free-free emission becomes optically thick and no longer follows a
$\nu^{-0.1}$ dependence.

$F^{856\micron}_{\rm ff}$ is listed in Table \ref{table2}, and represents the 
maximum level of free-free emission 
extrapolated to 856\,$\mu$m from Figure \ref{sed}.
The highest levels of centimeter emission were used to account for the free-free contributions
to result in the most conservative estimate of disk mass.
After accounting for the free-free contamination, we estimated $F_{\rm dust}$ for each 
source (see Table \ref{table2}); all 21 of the detected proplyds were detected in
thermal dust emission in excess of the free-free emission (see Figure \ref{sed}).

It is worth noting that we did not correct for background cloud emission 
as was done for the SMA observations.  The SMA synthesized beam size 
was relatively large in the Mann et al.~work (2$\farcs$5$\sim$1000\,AU) 
compared with the ALMA beam (0$\farcs$5$\sim$\,200\,AU).  
The emission probed by the ALMA data is sufficiently compact 
compared to the beam size that it is not likely to be 
contaminated severely by background emission.
Moreover, the many additional ALMA Cycle 0 baselines provide 
much better spatial frequency coverage, leading to greater image fidelity and
allowing a better separation of the disk emission from the background cloud.

Disk mass estimates and upper limits were then derived from the 
estimated $F_{\rm dust}$ values assuming the standard optically 
thin isothermal relationship \citep[e.g.,][]{beckwith90},
\begin{equation} \label{masseqn}
M_{\rm disk} = \frac{F_{\rm dust}d^2}{\kappa_{\nu}B_{\nu}(T)},
\end{equation}
where $d=400$\,pc is the distance to Orion \citep{sandstrom,menten,kraus07,
kraus09}, $\kappa_{\nu}=0.034$\,cm$^2$\,g$^{-1}$ is the \cite{beckwith90} dust 
grain opacity at 856\,$\mu$m with an implicit gas-to-dust mass ratio of 100:1,
and $B_{\nu}(T)$ is the Planck function.  We assume a typical disk temperature 
of 20\,K, as in previous disk surveys of Taurus and Ophiuchus by \citet{andrews05,
andrews07} and the ONC by \cite{mann09a,mann10}, for ease of comparison.  
Disk continuum emission can deviate from the optically thin limit if the column densities 
are especially large, an issue that was discussed in detail by \citet{andrews05}.  
The typical disk structure that could produce the observed 856\,$\micron$ flux densities in Orion
would imply the brighter disks could have up to $\sim$10-25\% of their emission being optically
thick (see Figure 20 of \citealt{andrews05}), resulting in an underestimation of their disk masses.

An overall disk mass sensitivity for the survey was determined by measuring the 
fraction of sources that could be detected at $\geq 3\sigma$ as a function of 
$M_{\rm disk}$, depending on the varying levels of free-free emission and 
target locations within each field.  We find that the observed fields are 
100\%\ complete for $M_{\rm disk} \geq\,1.2\,M_{\rm jup}$ ($1\,M_{\rm jup} = 
9.5 \times 10^{-4}\,M_\odot$) and 50\%\ complete for $M_{\rm disk} \ge 
0.4\,M_{\rm jup}$.  For comparison, the SMA survey at 880\,$\mu$m was 100\%\ 
complete for $M_{\rm disk}\geq\,8.7\,M_{\rm jup}$ \citep{mann10},
$\sim$\,7$\times$ higher than the ALMA data.

We were able to determine disk masses for all 21 proplyd detections, as there 
was sufficient dust emission in excess of the free-free contamination (see 
Table \ref{table2}).
In addition to the 22 {\it HST}-identified proplyds surveyed, we observed 14
sources from the ACS Survey of the {\it HST} Treasury Program \citep{robberto,
ricci}, 10 sources from the CTIO/Blanco 4\,m near-infared survey by 
\citet{robberto10}, and 2 massive stars, $\theta^1$\,Ori C and $\theta^1$\,Ori F
(see Table \ref{table3}).  Only 2 of the non-proplyd sources were detected, including
a newly discovered disk around 113-438 and the recovery of the disk around 
253-1536b, which was originally discovered through SMA imaging \citep{mann09b}.
The ALMA observations place stringent (3\,$\sigma$) upper 
limits of 0.1--0.6\,$M_{\rm jup}$ on the disk masses for the undetected targets
(see Table \ref{table3}).  These disks, if they exist, 
must be not only low in mass, but are likely smaller than $\sim$0\farcs15 
($\sim$60\,AU; \citealt{bally00,vicente}) to be unseen 
in the {\it HST} images.  No dust emission was 
detected toward the massive stars in this survey, $\theta^1$\,Ori C (spectral 
type O6) and $\theta^1$\,Ori F (spectral type B8).  In computing upper limits 
of $\sim$0.12\,$M_{\rm jup}$ for these targets, we adopted a higher dust 
temperature of 40\,K \citep[see][]{beuther02,Sridharan02}.  
The disk-to-stellar mass ratio for these massive stars
is $\lesssim$ 0.12\,$M_{\rm jup}$/40\,$M_{\sun}$ $\sim$ 3 $\times$10$^{-6}$, 
significantly lower than the typical range of 10$^{-1}$ to 10$^{-3}$ for 
T Tauri and HAeBe stars, implying that either massive stars do not form 
with disks or that their disks have much shorter lifetimes \citep{williams11}.

The giant silhouette disk 114--426 was detected for the first time at 
submillimeter wavelengths with these ALMA observations.  This disk has been one 
of the most puzzling objects in Orion, since it is the largest and most 
prominent optical disk in the entire ONC, but it had never been detected at 
long wavelengths \citep{bally98b,eisner06,eisner08,mann10}.  
An 1100\,AU disk seen nearly edge-on in {\it HST} images 
(see Figure \ref{fig2b}), 114--426 is found to have a surprisingly low flux of 
7\,mJy, over an order of magnitude less than the other giant silhouette 
disk in Orion, 216-0939 ($\sim$\,95\,mJy; see Table \ref{table2}).
The nature of this interesting disk will be the subject of a 
forthcoming article from our team (J. Bally et al., in preparation).

\subsection{Disk Masses and Distance from $\theta^1$\,Ori C}
Figure \ref{fig3} shows the disk masses (and flux densities at 856$\micron$) 
in the ONC as a function of their projected 
distance from the massive star $\theta^1$\,Ori C, including the previous SMA 
results from \citet{mann10} to fill in the intermediate distances not yet 
probed with ALMA.  The 3\,$\sigma$ upper limits for both surveys are indicated 
as grey arrows.  All of the known proplyds in the central field, i.e., within 9\arcsec\ 
of $\theta^1$\,Ori C ($\sim$0.02\,pc), are included.  This plot reveals clearly 
that disk masses tend to be substantially lower when they are located closer to 
$\theta^1$\,Ori C.  
Using the correlation tests described by \citet{isobe}, including the Cox Hazard
Model, Generalized Kendall's Tau, and Spearman's Rho tests,
that make use of the combined, censored dataset quantitatively confirm this trend, where 
the probability of no correlation between disk mass and projected distance to
$\theta^1$\,Ori C is $<$10$^{-4}$ for all 3 tests.  This provides a strong confirmation of the 
same relationship that was tentatively noted in the previous SMA survey data, 
particularly for the most massive disks \citep{mann10}.

\section{Discussion}\label{disc}
We observed the 856\,$\mu$m continuum emission toward 48 young stars in the 
Orion Nebula cluster using ALMA in Cycle 0, including 22 {\it HST}-identified 
proplyds.  With an overall 3$\sigma$ survey sensitivity limit of $\sim$1.2\,$M_{\rm 
jup}$, we detected 23 disks ($\sim$48\%), including 9 that had not been 
detected previously.  
Aside from the disks around 253--1536b and 113-438 (see Table \ref{table3}),
these detections coincide with the 
optically discovered disks from {\it HST} observations, highlighting the 
sensitivity of the space telescope to ONC disks due to their contrast with the 
bright nebular background.  The giant silhouette disk 114--426 was detected for 
the first time, and has a low flux density of 7\,mJy, and an estimated mass of 3.9\,$M_{\rm jup}$ 
(this disk will be the subject 
of a forthcoming article; J. Bally et al., in preparation).  Using this ALMA 
survey and the results of previous observations with the SMA, 
we find clear, statistically significant evidence for a marked decrease in the 856$\micron$ disk luminosities
of the Orion proplyds that have smaller projected separations from the massive star $\theta^1$\,Ori C.
In the assumption that the emission is optically thin, and the dust temperature and 
opacity are the same for all the disks, this implies that the masses of the
Orion proplyds decrease for those disks located near $\theta^1$\,Ori C.

\begin{figure}[h]
\centering
\includegraphics[scale=0.7]{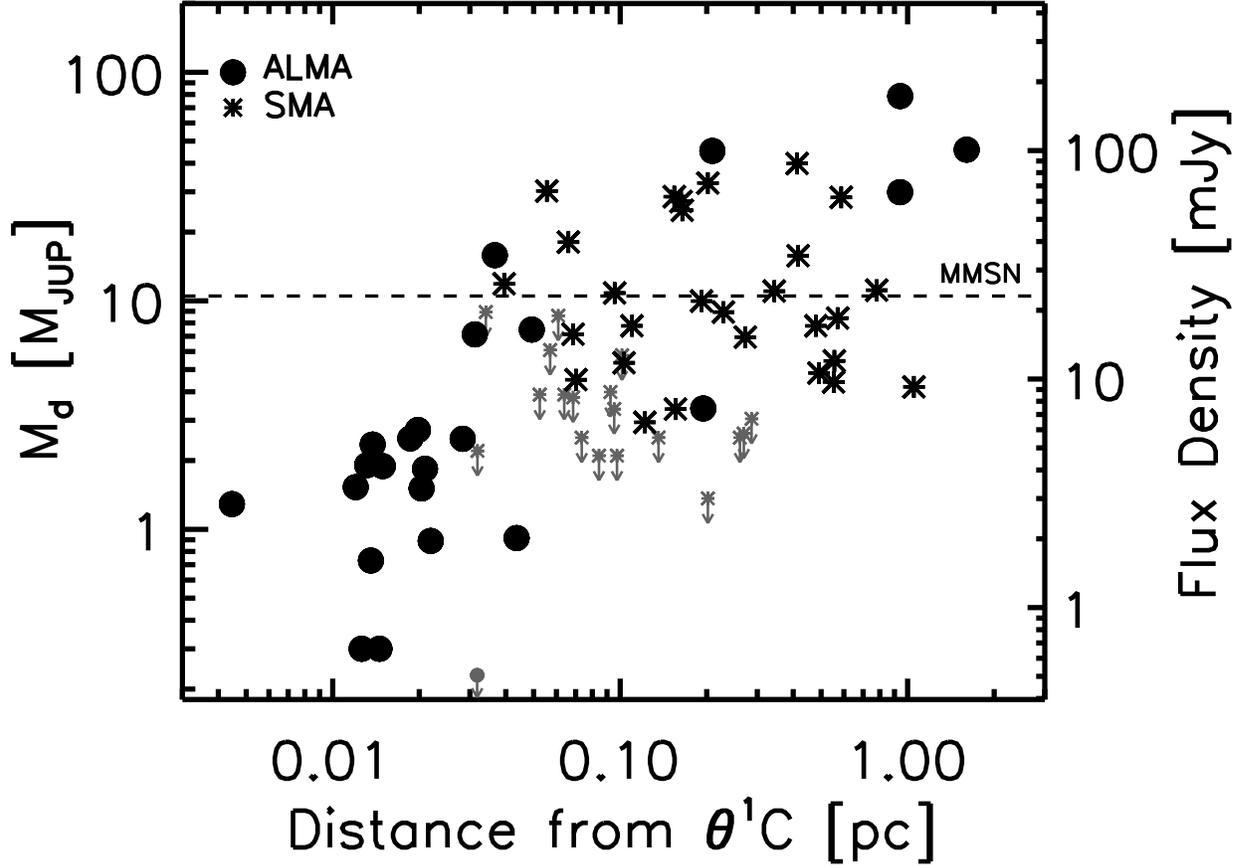}
\caption[Distance]{
Circumstellar disk masses plotted against their 
projected distances from the massive O-star, 
$\theta^1$\,Ori C.  Large black dots represent
ALMA detections, while stars represent SMA 
detections at 880$\micron$ for proplyds not yet 
observed with ALMA.  A grey dot and grey stars 
represent the 3\,$\sigma$ upper limits for the proplyds 
not detected in the observed fields.  In total, 70
{\em HST}-identified proplyds 
surveyed with both the ALMA and the SMA are plotted here.
The dashed line 
represents the MMSN value of 10$M_{\rm jup}$.
The observations expose the trend of decreasing disk masses
at smaller separations from $\theta^1$\,Ori C, particularly
within $\sim$\,0.03\,pc of the O-star, where there is a
lack of disks more massive than $\sim$\,3\,M$_{jup}$.
}
\label{fig3}
\end{figure}

The origins of that latter relationship could potentially be due to projection 
artifacts, initial conditions, and/or real evolutionary effects.  
The true separations between the proplyds and $\theta^1$\,Ori C are not 
known, but one can make a probabilistic argument that relates the projected 
separations to the true ones for an assumed distribution of orbital 
eccentricities around the ONC center of mass \citep[cf.][]{torres99}.  For a 
uniform eccentricity distribution, the projected and true separations should be 
commensurate within a factor of $\sim$2; for a steeper eccentricity 
distribution, the projected separations represent a more biased tracer of the 
true values and could be considered lower limits.  Such shifts in the abscissae 
of Figure \ref{fig3} would not explain the lack of disks around the targets with very close 
projected separations from $\theta^1$\,Ori C, nor do they seem likely to be 
large enough to erase the overall trend (although they indeed may adjust the 
basic shape). An intrinsic correlation between the masses of disks and their 
stellar hosts \citep[as found in the Taurus region by][]{andrews13} could 
account for the observed trend in Figure \ref{fig3} if the least massive stars are 
preferentially located near $\theta^1$\,Ori C.  Unfortunately, the nature of 
the Orion proplyds makes a direct determination of their stellar masses 
exceedingly difficult.  \citet{hillenbrand98} argued that stellar mass 
segregation in the ONC works in the opposite sense, with the highest mass stars 
($\ge$5\,$M_{\odot}$) concentrated toward the cluster center.  
If that were the case, we should have identified an anti-correlation between 
disk mass and distance from $\theta^1$\,Ori C, which is clearly not observed.  
High optical depths could be responsible for such a correlation, if the disks
located near $\theta^1$\,Ori C are smaller than the distant disks, and most of
their emission comes from optically thick regions.
However, no correlation has been observed between disk size and distance
from $\theta^1$\,Ori C \citep{vicente}.
Furthermore, a small ($\sim$\,50\,AU; the resolution of {\it HST}), 
completely optically thick disk would be detectable
by our sensitive ALMA observations, with a flux density of $\sim$\,55\,mJy
if viewed face-on, and an order of magnitude lower, $\sim$\,5.5\,mJy, 
if viewed nearly edge-on, suggesting the submillimeter wave optical depths
are not responsible for the observed correlation.

Instead, the evidence suggests that an externally driven disk evolution factor 
is likely responsible for the behavior in Figure \ref{fig3}.  Tidal stripping by stellar 
encounters is not only too inefficient for substantial disk destruction in the 
ONC \citep{scally01,hollenbach00}, but, as \citet{mann09b} argued, the conditions required 
for disk-disk interactions to deplete disk masses \citep[e.g.,][]{olczak} also 
implicitly involve very high photoevaporation mass-loss rates.  Overall, the 
data suggest that photoevaporative mass-loss driven by the ultraviolet 
radiation from $\theta^1$\,Ori C is the most dominant process responsible for 
the observed relationship.

Theoretical models of disk photoevaporation indeed predict mass-loss rates that 
decrease with distance from the irradiation source \citep{johnstone98,storzer,
richling00,scally01,matsuyama,adams04}.  These models suggest that only 
low-mass ($\la$ few $M_{\rm jup}$) disks should exist within 
$\sim$0.01--0.03\,pc of $\theta^1$\,Ori C because of the strong extreme-UV 
(EUV) irradiation at those distances \citep{johnstone98,storzer,adams04}.  At 
larger separations, $\sim$0.03--0.3\,pc, less energetic far-UV photons dominate 
the radiation field, resulting in lower mass-loss rates and thereby preserving 
more massive disks for up to a few Myr \citep[e.g.,][]{adams04}.

This predicted behavior is consistent with the observations in the context of 
Figure \ref{fig3}.  There is a clear lack of massive disks ($\gtrsim$\,3\,$M_{\rm jup}$) 
within 0.03\,pc of $\theta^1$\,Ori C where EUV irradiation dominates, whereas 
we find a wide range of disk masses (similar to what is found in low-mass star 
formation regions) at larger projected separations in the less destructive 
FUV-dominated regime.  
Accordingly, the potential to form a planetary system like our own in the 
EUV-dominated region of the ONC seems unlikely, given the substantially 
depleted disk masses there.  If these nearby disks have not formed planets already,
they may be out of luck unless 
dust grains have grown very large in these disks, 
to sizes not probed by submillimeter wavelength observations.
Resolved, multi-wavelength observations of the Orion proplyds are
required to investigate how far planet formation has already progressed in these young disks.
It is interesting to note, however, that the 
fraction of disks with masses that exceed the nominal ``Minimum Mass Solar 
Nebula" model \citep[$\sim$10\,$M_{\rm jup}$;][]{weidenschilling} in the more 
distant FUV-dominated region of the ONC is essentially the same as that found 
in the low-mass star formation environment of Taurus 
\citep[$\sim$10\%;][]{andrews13}.\footnote{Although it is worth noting that 
there are still strong selection effects at play in the currently incomplete 
ONC disk mass census that will need to be revisited when forthcoming ALMA 
datasets become available.}  
Overall, these observations support the idea that 
the strength of the local EUV irradiation field has profound environmental 
consequences on the potential for giant planet formation in the centers of 
massive star-forming regions.  

In ALMA Cycle 1, we expect to observe the disks around 300 stars in the ONC,
including 160 {\em HST}-identified proplyds.  This larger scale study will 
allow us to survey disks across a range of distances out to 1.6\,pc
from $\theta^1$\,Ori\,C, to probe different conditions in this massive star 
forming environment and uncover the overall disk fraction and the potential 
for forming planetary systems like our own.

\acknowledgments This paper makes use of the following ALMA data: 
ADS/JAO.ALMA\#2011.0.00028.S.  ALMA is a partnership of ESO (representing its 
member states), NSF (USA) and NINS (Japan), together with NRC (Canada) and NSC 
and ASIAA (Taiwan), in cooperation with the Republic of Chile.  The Joint ALMA 
Observatory is operated by ESO, AUI/NRAO and NAOJ.  The National Radio 
Astronomy Observatory is a facility of the National Science Foundation operated 
under cooperative agreement by Associated Universities, Inc.


\clearpage

\clearpage
\begin{deluxetable}{ccccccc}
\tablecolumns{7}
\tablewidth{0pc}
\tabletypesize{\scriptsize} 
\tablecaption{Summary of the ALMA Early Science Observations \label{table1}}
\tablehead{  
		\colhead{Field} & \colhead{$\alpha$ (J2000)} &
             	\colhead{$\delta$ (J2000)} & \colhead{Integration Time}  & 
		\colhead{1\,$\sigma$ rms} & 
		\colhead{Beam} & 
	        \colhead{PA}   \\
              
                \colhead{} & \colhead{[h m s]} & \colhead{[deg m s]} &
                \colhead{[sec]}  &  \colhead{[mJy beam$^{-1}$]}  & 
                \colhead{[$\arcsec$]} & 
		\colhead{[$\arcdeg$]}    
}
\startdata
1  &    05:35:16.30 &  -05:23:22.40  &  2600  & 0.24 &  0.50$\times$0.44 &    84 \\
2  &    05:35:17.35 &  -05:23:38.50  &  1300  & 0.24 &  0.51$\times$0.46 &    86  \\
3  &    05:35:11.40 &  -05:24:24.00  &  1300  & 0.58\footnotemark{} &  0.51$\times$0.46 &    88  \\
4  &    05:35:25.30 &  -05:15:35.50  &  1300  & 1.15 &  0.51$\times$0.46 &    89  \\
5  &    05:35:21.50 &  -05:09:42.00  &  1300  & 0.41 &  0.51$\times$0.46 &    89  \\

\enddata
\end{deluxetable}
\footnotetext{The rms value of Field 3 is itself uncertain by a factor of $\sim$2 due to the presence of extensive low-level emission in that Field.}

\clearpage
\begin{deluxetable}{clllrrrrrrl}
\tablecolumns{11}
\tablewidth{0pc}
\tabletypesize{\scriptsize} 
\tablecaption{22 Orion Proplyds Targeted in ALMA Early Science Observations \label{table2}}
\tablehead{  
		\colhead{Field} & \colhead{Proplyd} & \colhead{$\alpha$ (J2000)} &
             	\colhead{$\delta$ (J2000)} & 
		\colhead{F$_{856\,\micron}$}  & 
		\colhead{F$^{856\,\micron}_{\rm ff}$} & 
		\colhead{F$_{\rm dust}$} & 
	        \colhead{M$_{\rm disk}$}  &  \colhead{d [$\theta^1$C]}  & 
	        \colhead{Maj, Min, PA} &  
	        \colhead{Notes}  \\
              
                \colhead{} & \colhead{Name} & 
                \colhead{[h m s]} & \colhead{[deg m s]} &
        		\colhead{[mJy]} & 
		\colhead{[mJy]} & \colhead{[mJy]} & 
		\colhead{[Mjup]}  & \colhead{[pc]}  & \colhead{[AU, AU, deg]}
  
}
\startdata
1 & 157-323  &  5:35:15.74 & -5:23:22.49 &   2.8 $\pm$0.3  & 1.0 $\pm$0.1 &1.8$\pm$0.2  &  0.89 $\pm$0.12  & 0.022 &  ...,      ...,      ...       &      \\
1 & 158-323  &  5:35:15.84 & -5:23:22.47 &  15.6 $\pm$0.3 & 10.4 $\pm$0.9  & 5.1$\pm$0.9  &  2.50 $\pm$0.46  & 0.019 & 113, 22, +55 &   	\\
1 & 158-326  &  5:35:15.85 & -5:23:25.56 &  10.6 $\pm$0.3 &  5.0 $\pm$1.6   & 5.6$\pm$1.5   & 2.72 $\pm$0.74 & 0.020  & 144, 66, +57 &\\ 
1 & 158-327  &  5:35:15.79 & -5:23:26.56 &  22.9 $\pm$0.4 & 19.1 $\pm$1.4  & 3.8$\pm$1.4  & 1.84 $\pm$0.70 & 0.021  &147, 96, +45 & \\  
1 & 161-314  &  5:35:16.11 & -5:23:14.06 &   3.1 $\pm$0.6  &   0                    &  3.1$\pm$0.6   & 1.51 $\pm$0.29 & 0.020 &272, 211, +4 & new \\  
1 & 161-324  &  5:35:16.07 & -5:23:24.37 &   5.3 $\pm$0.3 &  4.7 $\pm$0.5   &  0.6$\pm$0.5   & 0.30 $\pm$0.27 & 0.013 & 103, 36, +83 & new \\
1 & 161-328  &  5:35:16.08 & -5:23:27.80 &   6.1 $\pm$0.3 &  2.2 $\pm$0.1 & 3.9$\pm$0.3  & 1.89 $\pm$0.17 & 0.015  & 193, 76, +67 &\\      
1 & 163-317  &  5:35:16.29 & -5:23:16.55 &  12.5 $\pm$0.3 & 7.6 $\pm$0.1 &  4.9$\pm$0.3  & 2.35 $\pm$0.15 & 0.014 & 87, 68, +38 & new \\  
1 & 163-323  &  5:35:16.33 & -5:23:22.56 &   6.4 $\pm$0.3 &  3.7 $\pm$0.5    &  2.7$\pm$0.6  & 1.29 $\pm$0.29 & 0.004 &154, 130, +61 &  \\
1 & 166-316  &  5:35:16.62 & -5:23:16.12 &   3.5 $\pm$0.4 &  2.0 $\pm$0.4    &  1.5$\pm$0.6   & 0.73 $\pm$0.28 & 0.014 & 215, 128, +83 & new \\  
1 & 167-317  &  5:35:16.75 & -5:23:16.44 &  26.0 $\pm$0.4 & 25.4 $\pm$1.3  &  0.6$\pm$1.4   & 0.30 $\pm$0.66 & 0.015 & 121, 118, +64 & \\
1 & 168-328  &  5:35:16.77 & -5:23:28.05 &   7.1 $\pm$0.3  &  3.1 $\pm$0.4   &  4.0$\pm$0.5   & 1.91 $\pm$0.25 & 0.013 & 131, 92, -65  &\\     
1 & 168-326  &  5:35:16.85 & -5:23:26.22 &  23.5 $\pm$0.5 & 20.3 $\pm$2.4  &   3.2$\pm$2.5   & 1.53 $\pm$1.18& 0.012 & 222, 118, -42 & new \\ 
2 & 169-338  &  5 35 16.88 & -5 23 38.10 & $\leq$\,0.8        &  0.3 $\pm$0.0 & $\leq$\,0.5 & $\leq$\,0.23 & 0.032 &  ...,  ..., ... & \\   
2 & 170-337  &  5:35:16.98 & -5:23:37.05 &  23.7 $\pm$0.4 &  8.9 $\pm$2.2   &  14.8$\pm$2.2  &  7.13 $\pm$1.08 & 0.031 &86, 73, +54   &  \\  
2 & 171-334  &  5:35:17.07 & -5:23:34.04 &  9.0 $\pm$0.4   &   3.8 $\pm$0.9  &   5.2$\pm$1.0   &  2.49 $\pm$0.49 & 0.028 & ..., ...,  ...      & new  \\ 
2 & 171-340  &  5:35:17.06 & -5:23:39.77 &  33.0 $\pm$0.5 &  0.2 $\pm$0.0 &  32.8$\pm$0.9  & 15.83 $\pm$0.42 & 0.037 & 102, 50, +53  & \\
2 & 173-341  &  5:35:17.34 & -5:23:41.49 &   2.9 $\pm$0.9   &  1.0 $\pm$0.1 &   1.9$\pm$0.5   &  0.92 $\pm$0.22 & 0.044 & 574, 37, -54* &  new\\
2 & 177-341  &  5:35:17.68 & -5:23:40.98 &  26.4 $\pm$0.5  & 10.9 $\pm$0.9 &  15.5$\pm$0.9  &  7.48 $\pm$0.45 & 0.049 & 125, 100, +58 &\\   
3 & 114-426  &  5:35:11.32 & -5:24:26.52 &   7.0 $\pm$1.2   &   0                   &   7.0$\pm$1.2   &  3.38 $\pm$0.56 & 0.195 &  ...,  ...,  ...    &  new \\
4 & 216-0939 &  5:35:21.58 & -5:09:38.96 &  94.9 $\pm$1.6  &   0                  &  94.9$\pm$1.6  & 45.84 $\pm$0.77 & 1.605 & 525, 150, -7  & \\   
5 & 253-1536 &  5:35:25.30 & -5:15:35.40 & 162.9 $\pm$0.9 &   0                  & 162.9$\pm$0.9  & 78.66 $\pm$0.42 & 0.942 &268, 95, +72  &  \\
\enddata
\tablecomments{
Column 1: Field location.
Column 2: Proplyd name.
Column 3, 4: Phase center coordinates. 
Column 5: Integrated continuum flux density, corrected for ALMA primary beam attenuation, with 1$\sigma$ statistical error.
Column 6: Extrapolated contribution of free-free emission at $856 \,\mu$m using the highest centimeter flux (see Figure \ref{sed}).
Column 7: Derived dust continuum flux density from the disk.
Column 8 : Disk mass from ALMA observations (error does not include uncertainties in the flux scale of $\sim$\,10\%). 
Column 9: Projected distance from $\theta^1$\,Ori C.
Column 10: Disk size: deconvolved semi-major axis, semi-minor axis, position angle; ``..." means that the source is effectively a point source.
*Note, the disk size for 173-341 is quite uncertain, given its faintness.
Column 11: New detections of submillimeter disk emission.
}
\end{deluxetable}

\clearpage
\begin{deluxetable}{lcllccr}
\tablecolumns{7}
\tablewidth{0pc}
\tabletypesize{\scriptsize} 
\tablecaption{26 Stars Targeted in ALMA Early Science Observations \label{table3}}
\tablehead{  
		\colhead{Field} & \colhead{Proplyd} & \colhead{$\alpha$ (J2000)} &
             	\colhead{$\delta$ (J2000)} & \colhead{F$_{856\,\micron}$}  & 
		\colhead{M$_{\rm disk}$}  & \colhead{Notes} \\
              
                \colhead{} & \colhead{Name} & 
                                \colhead{[h m s]} & \colhead{[deg m s]} & 
                                \colhead{[mJy]} & 
		\colhead{[Mjup]} 
		 
}
\startdata
1 &   157-326 &    5:35:15.73 & -5:23:25.66 &  $<$\,0.68   &   $<$\,0.33    &  ACS 4210 \\  
1 &   158-318 &    5:35:15.81 & -5:23:17.51 &  $<$\,0.69   &   $<$\,0.33    &  ACS 4246 \\  
1 &   160-323 &    5:35:15.97 & -5:23:22.74 &  $<$\,0.61   &   $<$\,0.29    &  ROB 20402 \\
1 &   161-323 &    5:35:16.10 & -5:23:23.20 &  $<$\,0.58   &   $<$\,0.28    &  ROB 2489  \\
1 &   162-319 &    5:35:16.24 & -5:23:19.13 &  $<$\,0.60   &   $<$\,0.29    &  ACS 4357 \\ 
1 &   163-328 &    5:35:16.28 & -5:23:27.55 &  $<$\,0.60   &   $<$\,0.29    &  ACS 4383  \\
1 &   164-321 &    5:35:16.37 & -5:23:21.15 &  $<$\,0.58   &   $<$\,0.28    &  ROB 2197 \\
1 &   164-325 &    5:35:16.35 & -5:23:25.34 &  $<$\,0.58   &   $<$\,0.28    &  ROB 20386  \\
1 &$\theta^1$\,C & 5:35:16.47 & -5:23:22.91 &  $<$\,0.58 &   $<$\,0.11    &  O6-type star \\
1 &   165-320 &    5:35:16.50 & -5:23:19.76 &  $<$\,0.58   &   $<$\,0.28    &  ACS 4427   \\
1 &   167-329 &    5:35:16.66 & -5:23:28.89 &  $<$\,0.65   &   $<$\,0.32    &  ACS 4486 \\
1 &$\theta^1$\,F & 5:35:16.72 & -5:23:25.20 &  $<$\,0.63  &   $<$\,0.12    &  B8-type star \\
2 &   168-342 &    5:35:16.84 & -5:23:42.28 &  $<$\,0.67   &   $<$\,0.32    &  ROB 3063 \\
2 &   173-337 &    5:35:17.28 & -5:23:37.20 &  $<$\,0.57   &   $<$\,0.28    &  ACS 4647,  \\
2 &   174-342 &    5:35:17.41 & -5:23:41.84 &  $<$\,0.58   &   $<$\,0.28    &  ACS 4696 \\
2 &   178-344 &    5:35:17.79 & -5:23:44.24 &  $<$\,0.67   &   $<$\,0.32    &  ACS 4831 \\
2 &   178-343 &    5:35:17.78 & -5:23:42.63 &  $<$\,0.64   &   $<$\,0.31    &  ACS 4825 \\
2 &   178-342 &    5:35:17.77 & -5:23:42.49 &  $<$\,0.64   &   $<$\,0.31    &  ACS 4827  \\
3 &   117-421 &    5:35:11.65 & -5:24:21.40 &  $<$\,0.87   &   $<$\,0.42    &  ACS 3388 \\
3 &   114-416 &    5:35:11.27 & -5:24:16.46 &  $<$\,0.91   &   $<$\,0.44    &  ROB 4371 \\
3 &   114-423 &    5:35:11.44 & -5:24:23.27 &  $<$\,0.82   &   $<$\,0.39    &  ACS 3360  \\
3 &   111-436 &    5:35:11.14 & -5:24:36.36 &  $<$\,1.13   &   $<$\,0.55    &  ACS 3335 \\ 
3 &   113-438 &    5:35:11.32 & -5:24:38.22 & 93.79$\pm$0.31 & 45.30$\pm$0.15 & ROBb 18 \\ 
4 &  216-0950 &    5:35:21.58 & -5:09:49.74 &  $<$\,0.95   &   $<$\,0.46    &  ROB 338 \\
4 &  218-0945 &    5:35:21.77 & -5:09:45.30 &  $<$\,0.89   &   $<$\,0.43    &  ROB 340\\
5 & 253-1536b &    5:35:25.23 & -5:15:35.69 & 61.86$\pm$2.0 & 29.88$\pm$0.95 &  ROB 6341 \\
\enddata
\tablecomments{
Column 1: Field Location.
Column 2: Proplyd Name.
Column 3, 4: Phase Center Coordinates (from \citet{ricci}).
Column 5: 3$\sigma$ dust continuum flux density upper limit.
Column 6: Disk mass upper limit.
Column 7: Notes: ACS sources from \citet{robberto}, ROB sources from \cite{robberto10},
ROBb source from \citet{robberto05}.
}
\end{deluxetable}

\clearpage
\bibliography{bib_rm}

\begin{thebibliography}{51}
\expandafter\ifx\csname natexlab\endcsname\relax\def\natexlab#1{#1}\fi

\bibitem[{{Adams} {et~al.}(2004){Adams}, {Hollenbach}, {Laughlin}, \&
  {Gorti}}]{adams04}
{Adams}, F.~C., {Hollenbach}, D., {Laughlin}, G., \& {Gorti}, U. 2004, \apj,
  611, 360

\bibitem[{{Andrews} {et~al.}(2013){Andrews}, {Rosenfeld}, {Kraus}, \&
  {Wilner}}]{andrews13}
{Andrews}, S.~M., {Rosenfeld}, K.~A., {Kraus}, A.~L., \& {Wilner}, D.~J. 2013,
  \apj, 771, 129

\bibitem[{{Andrews} \& {Williams}(2005)}]{andrews05}
{Andrews}, S.~M. \& {Williams}, J.~P. 2005, \apj, 631, 1134

\bibitem[{{Andrews} \& {Williams}(2007)}]{andrews07}
---. 2007, \apj, 671, 1800

\bibitem[{{Bally} {et~al.}(2000){Bally}, {O'Dell}, \& {McCaughrean}}]{bally00}
{Bally}, J., {O'Dell}, C.~R., \& {McCaughrean}, M.~J. 2000, \aj, 119, 2919

\bibitem[{{Bally} {et~al.}(1998{\natexlab{a}}){Bally}, {Sutherland}, {Devine},
  \& {Johnstone}}]{bally98}
{Bally}, J., {Sutherland}, R.~S., {Devine}, D., \& {Johnstone}, D.
  1998{\natexlab{a}}, \aj, 116, 293

\bibitem[{{Bally} {et~al.}(1998{\natexlab{b}}){Bally}, {Testi}, {Sargent}, \&
  {Carlstrom}}]{bally98b}
{Bally}, J., {Testi}, L., {Sargent}, A., \& {Carlstrom}, J. 1998{\natexlab{b}},
  \aj, 116, 854

\bibitem[{{Beckwith} {et~al.}(1990){Beckwith}, {Sargent}, {Chini}, \&
  {Guesten}}]{beckwith90}
{Beckwith}, S.~V.~W., {Sargent}, A.~I., {Chini}, R.~S., \& {Guesten}, R. 1990,
  \aj, 99, 924

\bibitem[{{Beuther} {et~al.}(2002){Beuther}, {Schilke}, {Menten}, {Motte},
  {Sridharan}, \& {Wyrowski}}]{beuther02}
{Beuther}, H., {Schilke}, P., {Menten}, K.~M., {Motte}, F., {Sridharan}, T.~K.,
  \& {Wyrowski}, F. 2002, \apj, 566, 945

\bibitem[{{Churchwell} {et~al.}(1987){Churchwell}, {Felli}, {Wood}, \&
  {Massi}}]{churchwell}
{Churchwell}, E., {Felli}, M., {Wood}, D.~O.~S., \& {Massi}, M. 1987, \apj,
  321, 516

\bibitem[{{Da Rio} {et~al.}(2009){Da Rio}, {Robberto}, {Soderblom}, {Panagia},
  {Hillenbrand}, {Palla}, \& {Stassun}}]{dario}
{Da Rio}, N., {Robberto}, M., {Soderblom}, D.~R., {Panagia}, N., {Hillenbrand},
  L.~A., {Palla}, F., \& {Stassun}, K. 2009, \apjs, 183, 261

\bibitem[{{Eisner} \& {Carpenter}(2006)}]{eisner06}
{Eisner}, J.~A. \& {Carpenter}, J.~M. 2006, \apj, 641, 1162

\bibitem[{{Eisner} {et~al.}(2008){Eisner}, {Plambeck}, {Carpenter}, {Corder},
  {Qi}, \& {Wilner}}]{eisner08}
{Eisner}, J.~A., {Plambeck}, R.~L., {Carpenter}, J.~M., {Corder}, S.~A., {Qi},
  C., \& {Wilner}, D. 2008, \apj, 683, 304

\bibitem[{{Felli} {et~al.}(1993{\natexlab{a}}){Felli}, {Churchwell}, {Wilson},
  \& {Taylor}}]{felli}
{Felli}, M., {Churchwell}, E., {Wilson}, T.~L., \& {Taylor}, G.~B.
  1993{\natexlab{a}}, \aaps, 98, 137

\bibitem[{{Felli} {et~al.}(1993{\natexlab{b}}){Felli}, {Taylor}, {Catarzi},
  {Churchwell}, \& {Kurtz}}]{felli93}
{Felli}, M., {Taylor}, G.~B., {Catarzi}, M., {Churchwell}, E., \& {Kurtz}, S.
  1993{\natexlab{b}}, \aaps, 101, 127

\bibitem[{{Garay} {et~al.}(1987){Garay}, {Moran}, \& {Reid}}]{garay}
{Garay}, G., {Moran}, J.~M., \& {Reid}, M.~J. 1987, \apj, 314, 535

\bibitem[{{Henney} \& {O'Dell}(1999)}]{henney}
{Henney}, W.~J. \& {O'Dell}, C.~R. 1999, \aj, 118, 2350

\bibitem[{{Hillenbrand} \& {Hartmann}(1998)}]{hillenbrand98}
{Hillenbrand}, L.~A. \& {Hartmann}, L.~W. 1998, \apj, 492, 540

\bibitem[{{Hollenbach} {et~al.}(2000){Hollenbach}, {Yorke}, \&
  {Johnstone}}]{hollenbach00}
{Hollenbach}, D.~J., {Yorke}, H.~W., \& {Johnstone}, D. 2000, Protostars and
  Planets IV, 401

\bibitem[{{Hubickyj} {et~al.}(2005){Hubickyj}, {Bodenheimer}, \&
  {Lissauer}}]{hubickyj05}
{Hubickyj}, O., {Bodenheimer}, P., \& {Lissauer}, J.~J. 2005, Icarus, 179, 415

\bibitem[{{Isobe} {et~al.}(1986){Isobe}, {Feigelson}, \& {Nelson}}]{isobe}
{Isobe}, T., {Feigelson}, E.~D., \& {Nelson}, P.~I. 1986, \apj, 306, 490

\bibitem[{{Johnstone} {et~al.}(1998){Johnstone}, {Hollenbach}, \&
  {Bally}}]{johnstone98}
{Johnstone}, D., {Hollenbach}, D., \& {Bally}, J. 1998, \apj, 499, 758

\bibitem[{{Kraus} {et~al.}(2007){Kraus}, {Balega}, {Berger}, {Hofmann},
  {Millan-Gabet}, {Monnier}, {Ohnaka}, {Pedretti}, {Preibisch}, {Schertl},
  {Schloerb}, {Traub}, \& {Weigelt}}]{kraus07}
{Kraus}, S., {Balega}, Y.~Y., {Berger}, J., {Hofmann}, K., {Millan-Gabet}, R.,
  {Monnier}, J.~D., {Ohnaka}, K., {Pedretti}, E., {Preibisch}, T., {Schertl},
  D., {Schloerb}, F.~P., {Traub}, W.~A., \& {Weigelt}, G. 2007, \aap, 466, 649

\bibitem[{{Kraus} {et~al.}(2009){Kraus}, {Weigelt}, {Balega}, {Docobo},
  {Hofmann}, {Preibisch}, {Schertl}, {Tamazian}, {Driebe}, {Ohnaka}, {Petrov},
  {Sch{\"o}ller}, \& {Smith}}]{kraus09}
{Kraus}, S., {Weigelt}, G., {Balega}, Y.~Y., {Docobo}, J.~A., {Hofmann}, K.,
  {Preibisch}, T., {Schertl}, D., {Tamazian}, V.~S., {Driebe}, T., {Ohnaka},
  K., {Petrov}, R., {Sch{\"o}ller}, M., \& {Smith}, M. 2009, \aap, 497, 195

\bibitem[{{Lada}(1998)}]{elada98}
{Lada}, E.~A. 1998, in Astronomical Society of the Pacific Conference Series,
  Vol. 148, Origins, ed. C.~E. {Woodward}, J.~M. {Shull}, \& H.~A. {Thronson},
  Jr., 198

\bibitem[{{Mann} \& {Williams}(2009{\natexlab{a}})}]{mann09a}
{Mann}, R.~K. \& {Williams}, J.~P. 2009{\natexlab{a}}, \apjl, 699, L55

\bibitem[{{Mann} \& {Williams}(2009{\natexlab{b}})}]{mann09b}
---. 2009{\natexlab{b}}, \apjl, 694, L36

\bibitem[{{Mann} \& {Williams}(2010)}]{mann10}
---. 2010, \apj, 725, 430

\bibitem[{{Matsuyama} {et~al.}(2003){Matsuyama}, {Johnstone}, \&
  {Hartmann}}]{matsuyama}
{Matsuyama}, I., {Johnstone}, D., \& {Hartmann}, L. 2003, \apj, 582, 893

\bibitem[{{McCullough} {et~al.}(1995){McCullough}, {Fugate}, {Christou},
  {Ellerbroek}, {Higgins}, {Spinhirne}, {Cleis}, \& {Moroney}}]{mccullough}
{McCullough}, P.~R., {Fugate}, R.~Q., {Christou}, J.~C., {Ellerbroek}, B.~L.,
  {Higgins}, C.~H., {Spinhirne}, J.~M., {Cleis}, R.~A., \& {Moroney}, J.~F.
  1995, \apj, 438, 394

\bibitem[{{Menten} {et~al.}(2007){Menten}, {Reid}, {Forbrich}, \&
  {Brunthaler}}]{menten}
{Menten}, K.~M., {Reid}, M.~J., {Forbrich}, J., \& {Brunthaler}, A. 2007, \aap,
  474, 515

\bibitem[{{Mundy} {et~al.}(1995){Mundy}, {Looney}, \& {Lada}}]{mundy}
{Mundy}, L.~G., {Looney}, L.~W., \& {Lada}, E.~A. 1995, \apjl, 452, L137+

\bibitem[{{O'Dell} \& {Wen}(1994)}]{odell94}
{O'Dell}, C.~R. \& {Wen}, Z. 1994, \apj, 436, 194

\bibitem[{{Olczak} {et~al.}(2006){Olczak}, {Pfalzner}, \& {Spurzem}}]{olczak}
{Olczak}, C., {Pfalzner}, S., \& {Spurzem}, R. 2006, \apj, 642, 1140

\bibitem[{{Reggiani} {et~al.}(2011){Reggiani}, {Robberto}, {Da Rio}, {Meyer},
  {Soderblom}, \& {Ricci}}]{reggiani11}
{Reggiani}, M., {Robberto}, M., {Da Rio}, N., {Meyer}, M.~R., {Soderblom},
  D.~R., \& {Ricci}, L. 2011, \aap, 534, A83

\bibitem[{{Ricci} {et~al.}(2008){Ricci}, {Robberto}, \& {Soderblom}}]{ricci}
{Ricci}, L., {Robberto}, M., \& {Soderblom}, D.~R. 2008, \aj, 136, 2136

\bibitem[{{Richling} \& {Yorke}(2000)}]{richling00}
{Richling}, S. \& {Yorke}, H.~W. 2000, \apj, 539, 258

\bibitem[{{Robberto} {et~al.}(2005){Robberto}, {Beckwith}, {Panagia}, {Patel},
  {Herbst}, {Ligori}, {Custo}, {Boccacci}, \& {Bertero}}]{robberto05}
{Robberto}, M., {Beckwith}, S.~V.~W., {Panagia}, N., {Patel}, S.~G., {Herbst},
  T.~M., {Ligori}, S., {Custo}, A., {Boccacci}, P., \& {Bertero}, M. 2005, \aj,
  129, 1534

\bibitem[{{Robberto} {et~al.}(2013){Robberto}, {Soderblom}, {Bergeron},
  {Kozhurina-Platais}, {Makidon}, {McCullough}, {McMaster}, {Panagia}, {Reid},
  {Levay}, {Frattare}, {Da Rio}, {Andersen}, {O'Dell}, {Stassun}, {Simon},
  {Feigelson}, {Stauffer}, {Meyer}, {Reggiani}, {Krist}, {Manara},
  {Romaniello}, {Hillenbrand}, {Ricci}, {Palla}, {Najita}, {Ananna},
  {Scandariato}, \& {Smith}}]{robberto}
{Robberto}, M., {Soderblom}, D.~R., {Bergeron}, E., {Kozhurina-Platais}, V.,
  {Makidon}, R.~B., {McCullough}, P.~R., {McMaster}, M., {Panagia}, N., {Reid},
  I.~N., {Levay}, Z., {Frattare}, L., {Da Rio}, N., {Andersen}, M., {O'Dell},
  C.~R., {Stassun}, K.~G., {Simon}, M., {Feigelson}, E.~D., {Stauffer}, J.~R.,
  {Meyer}, M., {Reggiani}, M., {Krist}, J., {Manara}, C.~F., {Romaniello}, M.,
  {Hillenbrand}, L.~A., {Ricci}, L., {Palla}, F., {Najita}, J.~R., {Ananna},
  T.~T., {Scandariato}, G., \& {Smith}, K. 2013, \apjs, 207, 10

\bibitem[{{Robberto} {et~al.}(2010){Robberto}, {Soderblom}, {Scandariato},
  {Smith}, {Da Rio}, {Pagano}, \& {Spezzi}}]{robberto10}
{Robberto}, M., {Soderblom}, D.~R., {Scandariato}, G., {Smith}, K., {Da Rio},
  N., {Pagano}, I., \& {Spezzi}, L. 2010, \aj, 139, 950

\bibitem[{{Sandstrom} {et~al.}(2007){Sandstrom}, {Peek}, {Bower}, {Bolatto}, \&
  {Plambeck}}]{sandstrom}
{Sandstrom}, K.~M., {Peek}, J.~E.~G., {Bower}, G.~C., {Bolatto}, A.~D., \&
  {Plambeck}, R.~L. 2007, \apj, 667, 1161

\bibitem[{{Scally} \& {Clarke}(2001)}]{scally01}
{Scally}, A. \& {Clarke}, C. 2001, \mnras, 325, 449

\bibitem[{{Smith} {et~al.}(2005){Smith}, {Bally}, {Licht}, \&
  {Walawender}}]{smith}
{Smith}, N., {Bally}, J., {Licht}, D., \& {Walawender}, J. 2005, \aj, 129, 382

\bibitem[{{Sridharan} {et~al.}(2002){Sridharan}, {Beuther}, {Schilke},
  {Menten}, \& {Wyrowski}}]{Sridharan02}
{Sridharan}, T.~K., {Beuther}, H., {Schilke}, P., {Menten}, K.~M., \&
  {Wyrowski}, F. 2002, \apj, 566, 931

\bibitem[{{St{\"o}rzer} \& {Hollenbach}(1999)}]{storzer}
{St{\"o}rzer}, H. \& {Hollenbach}, D. 1999, \apj, 515, 669

\bibitem[{{Torres}(1999)}]{torres99}
{Torres}, G. 1999, \pasp, 111, 169

\bibitem[{{Vicente} \& {Alves}(2005)}]{vicente}
{Vicente}, S.~M. \& {Alves}, J. 2005, \aap, 441, 195

\bibitem[{{Weidenschilling}(1977)}]{weidenschilling}
{Weidenschilling}, S.~J. 1977, \apss, 51, 153

\bibitem[{{Williams} {et~al.}(2005){Williams}, {Andrews}, \&
  {Wilner}}]{williams}
{Williams}, J.~P., {Andrews}, S.~M., \& {Wilner}, D.~J. 2005, \apj, 634, 495

\bibitem[{{Williams} \& {Cieza}(2011)}]{williams11}
{Williams}, J.~P. \& {Cieza}, L.~A. 2011, \araa, 49, 67

\bibitem[{{Zapata} {et~al.}(2004){Zapata}, {Rodr{\'{\i}}guez}, {Kurtz}, \&
  {O'Dell}}]{zapata}
{Zapata}, L.~A., {Rodr{\'{\i}}guez}, L.~F., {Kurtz}, S.~E., \& {O'Dell}, C.~R.
  2004, \aj, 127, 2252

\end{thebibliography}

\end{document}